\documentclass[onecolumn,superscriptaddress]{revtex4}

\usepackage{graphicx,float}
\usepackage{epsfig}
\usepackage{dcolumn}
\usepackage{ulem}
\usepackage{bm}
\usepackage{amsmath}
\usepackage{amssymb}
\usepackage{hyperref}
\hypersetup{colorlinks=true,linkcolor=blue,citecolor=blue,urlcolor=black}

\newcommand{\ket}[1]{\ensuremath{|\,{#1}\,\rangle}}

\begin{document}

\title{High-dimensional decoy-state quantum key distribution over 0.3 km of multicore telecommunication optical fibers}
\date{\today}

\author{G.~Ca\~nas}\thanks{These authors contributed equally to this work}
\affiliation{Departamento de F\'isica, Universidad de Concepci\'on, 160-C Concepci\'on, Chile}
\affiliation{Center for Optics and Photonics, Universidad de Concepci\'on, Concepci\'on, Chile}
\affiliation{MSI-Nucleus for Advanced Optics, Universidad de Concepci\'on, Concepci\'on, Chile}

\author{N.~Vera}\thanks{These authors contributed equally to this work}
\affiliation{Departamento de F\'isica, Universidad de Concepci\'on, 160-C Concepci\'on, Chile}
\affiliation{Center for Optics and Photonics, Universidad de Concepci\'on, Concepci\'on, Chile}
\affiliation{MSI-Nucleus for Advanced Optics, Universidad de Concepci\'on, Concepci\'on, Chile}

\author{J.~Cari\~ne}\thanks{These authors contributed equally to this work}
\affiliation{Center for Optics and Photonics, Universidad de Concepci\'on, Concepci\'on, Chile}
\affiliation{MSI-Nucleus for Advanced Optics, Universidad de Concepci\'on, Concepci\'on, Chile}
\affiliation{Departamento de Ingenier\'ia El\'ectrica, Universidad de Concepci\'on, 160-C Concepci\'on, Chile}

\author{P.~Gonz\'alez}
\affiliation{Departamento de F\'isica, Universidad de Concepci\'on, 160-C Concepci\'on, Chile}
\affiliation{Center for Optics and Photonics, Universidad de Concepci\'on, Concepci\'on, Chile}
\affiliation{MSI-Nucleus for Advanced Optics, Universidad de Concepci\'on, Concepci\'on, Chile}

\author{J.~Cardenas}
\affiliation{Center for Optics and Photonics, Universidad de Concepci\'on, Concepci\'on, Chile}
\affiliation{Departamento de Ingenier\'ia El\'ectrica, Universidad de Concepci\'on, 160-C Concepci\'on, Chile}

\author{P.~W.~R.~Connolly}
\altaffiliation[Current address: ]{Physics Department, David Brewster Building, Heriot-Watt University, Edinburgh EH14 4AS, Scotland}
\affiliation{Departamento de F\'isica, Universidad de Concepci\'on, 160-C Concepci\'on, Chile}
\affiliation{Center for Optics and Photonics, Universidad de Concepci\'on, Concepci\'on, Chile}
\affiliation{MSI-Nucleus for Advanced Optics, Universidad de Concepci\'on, Concepci\'on, Chile}

\author{A.~Przysiezna}
\affiliation{Institute of Theoretical Physics and Astrophysics, Faculty of Mathematics, Physics and Informatics, University of Gda\'nsk,
80-308 Gda\'nsk, Poland}

\author{E.~S.~G\'omez}
\affiliation{Departamento de F\'isica, Universidad de Concepci\'on, 160-C Concepci\'on, Chile}
\affiliation{Center for Optics and Photonics, Universidad de Concepci\'on, Concepci\'on, Chile}
\affiliation{MSI-Nucleus for Advanced Optics, Universidad de Concepci\'on, Concepci\'on, Chile}

\author{M.~Figueroa}
\affiliation{Center for Optics and Photonics, Universidad de Concepci\'on, Concepci\'on, Chile}
\affiliation{Departamento de Ingenier\'ia El\'ectrica, Universidad de Concepci\'on, 160-C Concepci\'on, Chile}

\author{G.~Vallone}
\affiliation{Dipartimento di Ingegneria dell'Informazione, Universit\`a degli Studi di Padova, Padova 35131, Italy}
\affiliation{Istituto di Fotonica e Nanotecnologie, CNR, Padova, Italy}

\author{P.~Villoresi}
\affiliation{Dipartimento di Ingegneria dell'Informazione, Universit\`a degli Studi di Padova, Padova 35131, Italy}
\affiliation{Istituto di Fotonica e Nanotecnologie, CNR, Padova, Italy}

\author{T.~Ferreira~da~Silva}
\affiliation{Optical Metrology Division, National Institute of Metrology, Quality and Technology,
 25250-020 Duque de Caxias, RJ, Brazil}

 \author{G.~B.~Xavier}
\affiliation{Center for Optics and Photonics, Universidad de Concepci\'on, Concepci\'on, Chile}
\affiliation{MSI-Nucleus for Advanced Optics, Universidad de Concepci\'on, Concepci\'on, Chile}
\affiliation{Departamento de Ingenier\'ia El\'ectrica, Universidad de Concepci\'on, 160-C Concepci\'on, Chile}

\author{G.~Lima}
\email{glima@udec.cl}
\affiliation{Departamento de F\'isica, Universidad de Concepci\'on, 160-C Concepci\'on, Chile}
\affiliation{Center for Optics and Photonics, Universidad de Concepci\'on, Concepci\'on, Chile}
\affiliation{MSI-Nucleus for Advanced Optics, Universidad de Concepci\'on, Concepci\'on, Chile}

\maketitle

%%%%%%%%%%%%%%%%%%%%%%%%%%%%%%%%%%%%%%%%%%%%%%%%%%%%%%%%%%%%%%%%

\textbf{Multiplexing is a strategy to augment the transmission capacity of a communication system. It consists of combining multiple signals over the same data channel and it has been very successful in classical communications. However, the use of enhanced channels has only reached limited practicality in quantum communications (QC) as it requires the complex manipulation of quantum systems of higher dimensions. Considerable effort is being made towards QC using high-dimensional quantum systems encoded into the transverse momentum of single photons but, so far, no approach has been proven to be fully compatible with the existing telecommunication infrastructure. Here, we overcome such a technological challenge and demonstrate a stable and secure high-dimensional decoy-state quantum key distribution session over a 0.3 km long multicore optical fiber. The high-dimensional quantum states are defined in terms of the multiple core modes available for the photon transmission over the fiber, and the decoy-state analysis demonstrates that our technique enables a positive secret key generation rate up to 25 km of fiber propagation. Finally, we show how our results build up towards a high-dimensional quantum network composed of free-space and fiber based links.}
\newpage

In an age defined by several technological breakthroughs, we are aware that our privacy will be threatened by the likely development of quantum computers. Yet, we are confident that countermeasures will be created allowing post-quantum cryptography \cite{Post-quantum}. One possibility is the use of quantum-resistant classical cryptographic algorithms that provides a patch-safe solution for private communication to everyday internet users. Unfortunately, however, this method falls short while considered for sensitive documents of big corporations as classical signals can be copied and stored to be decrypted decades ahead. In this context, quantum cryptography emerges as a necessary and complementary alternative for modern global secure communications, since the certifiable security provided by this technique can not be compromised after the communication has been performed \cite{Gisin_RMP_2002, Lo_natphoton_2014, Diamanti_arxiv_2016}. Thus, it provides the long-term privacy required in many cases.

Over the last decades we have witnessed the advances of telecommunication technologies by experiencing a huge increase on our capacity to send/download data. This has been vastly based on the development of new techniques to multiplex information in different degrees of freedom of light transmitted over an optical fiber, which have allowed their information capacity to be increased around tenfold every four years \cite{Richardson_natphoton_2013}. Analogously, in quantum communications, the use of high-dimensional quantum systems allows for more information to be transmitted between the communicating parties \cite{Cerf_PRL_2002}. Fortunately, it turns out that such complex quantum systems can be created by also exploring the degrees of freedom of faint light pulses (attenuated to the single-photon level), and therefore most of the multiplexing strategies developed for classical telecommunications are to some extent connected to the implementation of high-dimensional secure quantum communications. This hardware compatibility, considered together with the historical development of classical telecommunications that had to deal with an ever-growing internet traffic, shows that if quantum technologies are to emerge as an alternative solution for the post-quantum cryptography era, then it will rely on the use of high-dimensional quantum systems.

Even though experimental high-dimensional quantum cryptography is still at its infancy, secure communications based on the use of high-dimensional quantum systems encoded into the transverse momentum of single photons has been the subject of many recent experimental efforts \cite{Steve_PRL_2006, Howell_PRL_2007, Etcheverry_Scirep_2013, Mafu_PRA_2013, Boyd_NJP_2015}, and theoretical analyses \cite{Cerf_PRL_2002, Sheridan_PRA_2010, Bunandar_PRA_2015, Bao_PhysA_2016, Bao_Opex_2016, Niu_Arxiv_2016, Bradler_NJP_2016}. The motivation comes from the versatility provided by the fact that it can be used to define an infinite-dimensional Hilbert space in terms of the orbital angular momentum (OAM) of Laguerre-Gaussian single-photon modes \cite{Leach_PRL_2002}, or also in terms of the number of linear transverse modes available for the photon transmission \cite{Neves_PRL_2005}. OAM encoded quantum systems are suitable for communication over free-space links due to its resilience against perturbation effects caused by atmospheric turbulence \cite{Rodenburg_opex_2012}, while on the other hand, path encoded quantum states are suitable for communications systems based on waveguide integrated circuits \cite{Ciampini_LSA_2016}. However, all the implementations performed so far suffer of severe drawbacks. For instance, all of them have been limited to low bandwidth as the repetition rate lies at the range of kHz, and most important, no research proposed so far has accomplished a secure quantum communication session while propagating such quantum states over the already available telecommunication fiber based infrastructure, casting serious doubts about its viability for real world applications.

Here we take a major step overcoming this last technological challenge and demonstrate a secure high-dimensional quantum key distribution (HD-QKD) session, between two parties communicating over a 0.3 km long telecommunication optical fiber, whose security is guaranteed by resorting to the decoy-state method. Our technique is built upon newly developed multicore optical fibers, now used in classical telecommunications for space-division multiplexing \cite{Richardson_natphoton_2013}. By modifying the hardware at the input and output of a this 0.3 km long multicore fiber we are able to coherently propagate a quantum signal over its entire length, thus, allowing high-fidelity transmission of 4-dimensional quantum systems that are encoded into the four available core modes. Using a standard InGaAs gated single-photon detector with a detection efficiency of 6$\%$, operating at a repetition rate of 1~kHz and with a dark count probability of $2.25 \times 10^{-7}$, we obtain a secret key bit generation rate per pulse of $(4.31 \pm 1.19) \times 10^{-6}$. We prove the HD-QKD session to be highly stable maintaining an overall quantum bit error rate (QBER) of $(10.25 \pm 0.6) \%$ over more than 20 hours of continuous operation. The decoy-state analysis \cite{Hwang_PRL_2003, Lo_PRL_2005, Wang_PRL_2005} further shows that our technique enables a positive secret key bit generation probability up to 25 km of fiber propagation. Last, we clarify that our technique is fully compatible with research towards quantum communications based on OAM encoded high-dimensional quantum states. We show that a new interface device can be constructed that allows the efficient coupling of 4-dimensional OAM quantum states into multicore fibers. That is, the device allows for the efficient mapping of OAM-encoded to core-path-encoded quantum states, and vice-versa. The device is a modified version of previous reported OAM mode sorters \cite{Berkhout_PRL_2010}, which is now suitable for multicore fiber configurations. These results pave the way towards a high-dimensional quantum network composed of interconnected free-space and optical fiber links.

\section*{Results}

\subsubsection*{The BB84 QKD session and the decoy-state technique relevance}
\label{Results}

\begin{figure}[t]
\centerline{\includegraphics[width=0.7 \textwidth]{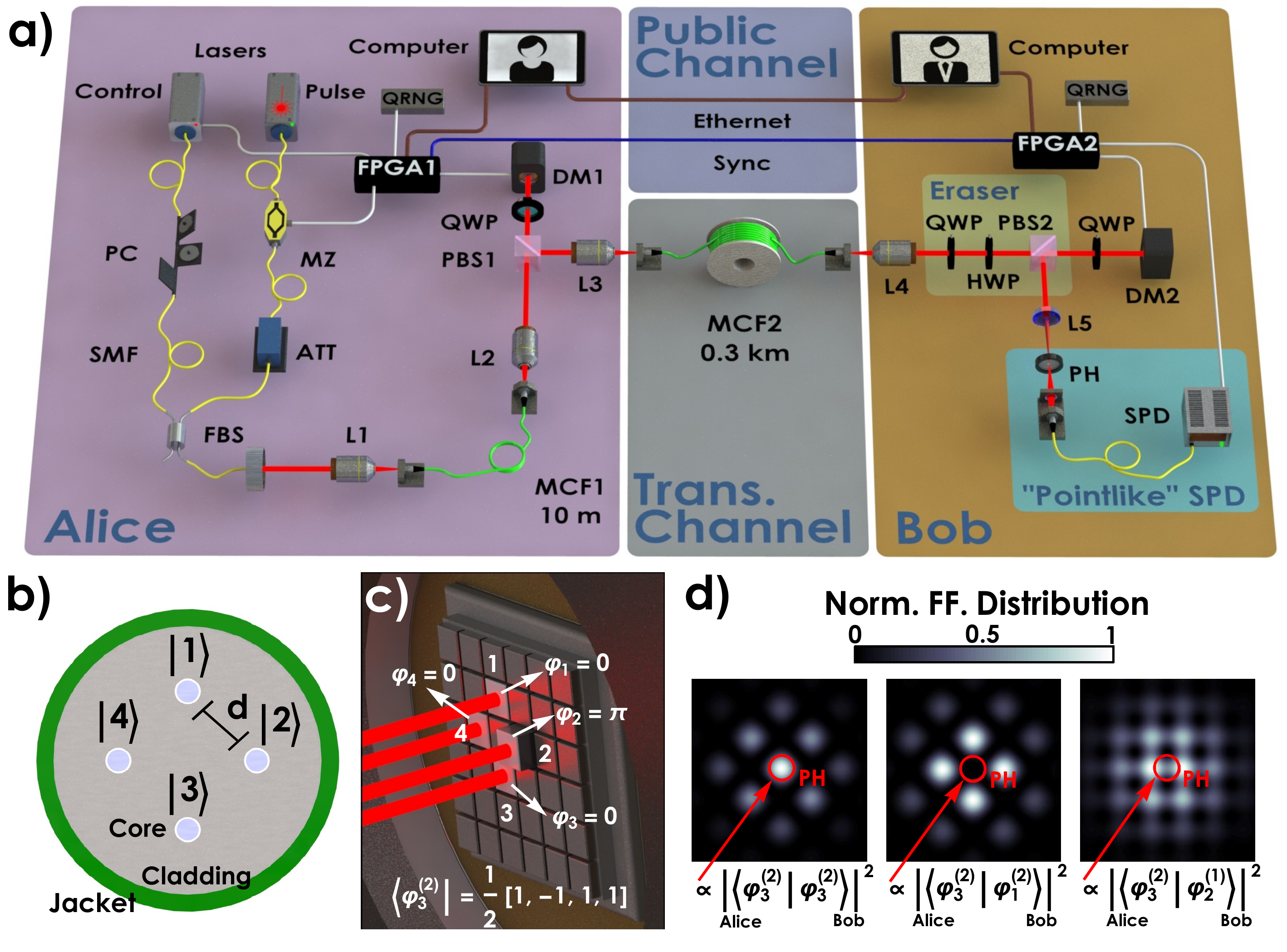}}
\caption{\textbf{Experimental setup.} a) In our scheme Alice employs a source of weak coherent states to encode the 4-dimensional BB84 QKD states using a deformable mirror (DM1). The single photons are sent to Bob through a 0.3 km long four-core multicore fiber. Bob employs a quantum eraser to get rid of any possible polarization-mode coupling during fiber propagation. He then uses an identical deformable mirror to Alice's (DM2) and a ``pointlike'' SPD, to implement his measurements. The QKD protocol is automatically executed using two FPGA electronic modules, fed with QRNGs. Finally each FPGA sends its results to a computer, which are used to determine the session's QBER. A control laser is used to periodically check whether Alice and Bob's referential frames are aligned, and the FPGAs also command this control procedure.  b) Schematic of the multicore fiber's cross-section. c) The deformable mirror is composed of a 6 $\times$ 6 mirror matrix (Boston micromachines). The light coming from each core from the MCF is mapped to an individual mirror. As an example, cores $\ket{1}$, $\ket{3}$ and $\ket{4}$ have a relative phase of 0 applied, while $\ket{2}$ has a $\pi$ relative phase-shift. d) Simulation of the FF distribution, with the pinhole area indicated by the red circle. The first case shows when Bob's projection is performed on the same state as the one Alice sent, both using the same MUB. It displays constructive interference through the pinhole. In the second case the pattern shows a situation where an orthogonal projection is used within the same MUB, leading to destructive interference and no detection. The final case happens when any projection is made using a different MUB within respect to Alice's. Then there is a 25\% probability that the photon will be detected.  ATT: Adjustable optical attenuator; DM: Deformable mirror; FBS: Fiber beamsplitter; HWP: Half-wave plate; L: Objective lens; MCF: Multicore fiber; MZ: Mach-Zehnder amplitude modulator; PBS: Polarizing beamsplitter; PH: Pinhole; QRNG: Quantum random number generator; QWP: Quarter-wave plate; SPD: Single-photon detector; SMF: Single-mode fiber.} \label{Fig1}
\end{figure}

By far the most widely used QKD protocol is BB84, which requires a prepare-and-measure scheme \cite{BB84}. The BB84 QKD session typically consists of Alice (the transmitter) randomly encoding bits of information onto single photons, and then sending them to Bob (the receiver) over an optical fiber or a free-space link. Alice's encoding procedure randomly chooses between two mutually unbiased bases (MUBs) for each bit to be sent, and Bob independently also randomly chooses states between the same two MUBs to perform a projective measurement on each photon. Alice and Bob then acknowledge through a classical authenticated channel which MUB was used, a procedure called basis reconciliation. They are now in possession of a shared string, which contains errors due to the transmission and detection processes. Alice and Bob then estimate the error rate on their string, apply error correction procedures on it and finally, in order to ensure that an eavesdropper (Eve) information gain on the transmitted bits is null, apply privacy amplification techniques \cite{Gisin_RMP_2002}.

One major practical problem in QKD implementations is the fact that ideal single-photon sources are not available, although there are recent promising alternatives \cite{Ding_PRL_2016}. Nevertheless the most practical source by far is an attenuated laser producing weak coherent states. The main issue when using these sources that deviate from ideal ones, is that Eve may perform the so-called ``photon-number splitting'' attack (PNS) on pulses that contain more than one photon \cite{Huttner_PRA_1995}. The solution to this problem is the decoy-state method \cite{Hwang_PRL_2003, Lo_PRL_2005, Wang_PRL_2005}, where Alice intentionally and randomly prepares states with different average photon numbers per pulse. During the classical reconciliation phase she tells Bob which pulses correspond to a particular photon number, and from the detection statistics for each type of pulse, they can estimate more precisely the fraction of detected single-photon pulses and determine whether an eavesdropper is present. Due to its relatively straightforward implementation, the decoy-state method has been used in most QKD experiments in recent years \cite{Dixon_Opex_2008, Zhao_PRL_2006, Peng_PRL_2007, Liu_Opex_2010, Silva_PRA_2013, Liu_PRL_2013, Tang_PRL_2014}. In our implementation we use the generalization of the decoy-state technique developed for the HD-QKD BB84 protocol \cite{Cerf_PRL_2002, Sheridan_PRA_2010}.

\subsubsection*{Single-photon source}

In our work we employ a heavily attenuated telecom distributed feedback laser, whose emission wavelength is 1546.32 nm, as our source of weak coherent states [Fig. \ref{Fig1}(a)]. The laser operates in continuous wave (CW) mode and is externally modulated by a Mach-Zehnder electro-optical modulator (MZ), generating 500~ps wide optical pulses. A calibrated optical attenuator (ATT) is used to set the desired average photon number per pulse, $\mu$, at Alice's output. Distinct QKD sessions have been implemented in our work, but the highest average photon number per pulsed adopted was $\mu=0.27$. In this case, the probability of having non-null pulses, i.e., of having pulses containing at least one photon is $P(\mu=0.27|n\geq1)=23.7\%$. Pulses containing only one photon are the vast majority of the non-null pulses generated ($\sim90\%$). The repetition frequency for the optical pulses is limited to 1~kHz due to restrictions on the preparation (measurement) of Alice (Bob) quantum states, as explained below. Last, note that since the period between consecutive pulses (1 ms) is much longer than the coherence time of the laser ($\sim 0.1 \mu$s), there is no need to employ active phase randomization of the pulses, avoiding potential security loopholes \cite{Zhao_APL_2007}.

\subsubsection*{Alice state generation}

The probabilistically generated single photons are then used at Alice's site to encode the required high-dimensional quantum states for the QKD session. For this purpose the attenuated pulses are initially coupled into a 10 m long Fibercore multicore fiber (MCF1), composed of four single-mode cores, by means of a $10 \times$ objective lens (L1) [See Fig.~\ref{Fig1}(a)-(b)]. The core mode field diameter is 8.5$\mu$m and the cores are separated by a distance $d=36.25~\mu$m to ensure that cross-talk effects are negligible. The input face of the fiber is positioned slightly out of the lens focal plane such that all cores of the fiber are equally illuminated. Thus, the probability amplitudes for the photon transmission by each core are equally weighted. Contrary to standard fiber arrays, the cores of multicore fibers lie within the same cladding, ensuring that random phase-fluctuations induced by thermal and mechanical stress are strongly suppressed. Therefore, the state of the single photons transmitted over the MCF1 can be written as a coherent superposition given by $|\Psi\rangle = \frac{1}{2}\sum_{1}^{4}e^{i\phi_l}|l\rangle$, where $|l\rangle$ denotes the state of the photon transmitted by the $l$th transverse core mode, and $\phi_l$ is the relative phase acquired during the propagation over the $l$th core. This is the fiducial state which is then used to prepare the required states for the 4-dimensional BB84 QKD session.

The 4-dimensional BB84 QKD states span two MUBs. The states are given explicitly in the Methods section and are denoted by $|\varphi_i^{(j)}\rangle$, where $i=1,2,3,4$ refers to the $i$th state of the $j$th MUB, with $j=1,2$. Alice state preparation is done by imaging the output face of the MCF1 onto a deformable mirror (DM1) by means of a second $10 \times$ objective lens (L2). The 10$\times$ magnification factor is intentionally chosen such that the image of each core is formed at different mirrors belonging to the DM1, as shown schematically in Fig.~\ref{Fig1}(c). Each mirror longitudinal position can be set individually. By defining different offset positions for the four mirrors, the residual phases $\phi_l$ are compensated and the first state $|\varphi_1^{(1)}\rangle$ prepared. The other QKD states are generated by calibrating in respect to the other longitudinal position of each mirror that corresponds to a relative phase-shift $\varphi_l = \pi$, also schematically shown in Fig.~\ref{Fig1}(c).

The single photons are then coupled back to a similar but 0.3~km long multicore fiber (MCF2) [resorting to a third $10 \times$ objective lens (L3)], comprising the transmission channel to Bob's station. Finally it is important to note that during Alice's preparation stage, polarizing optics are used to ensure that there is no coupling/entanglement between the polarization and the modes available for the photon transmission [See Fig.~\ref{Fig1}(a)]. Thus, ensuring that there is no state information from Alice's available in the polarization degree of freedom of the photons sent to Bob, which could be exploited by an eavesdropper.

\subsubsection*{Bob state detection}

\begin{table}[t]
\begin{center}
\begin{tabular}{|c|c|c|}
\hline
PH diameter ($\mu m$) & Loss ($dB$) & QBER (\%)\\
\hline
$25$ & $-19.19$ & $2.3$\\
\hline
$50$ & $-13.57$ & $8.8$\\
\hline
$75$ & $-10.58$ & $18.8$\\
\hline
$100$ & $-8.82$ & $30.7$\\
\hline
\end{tabular}
\caption{\textbf{PH diamenter $\times$ losses $\times$ QBER.} Dependence of the losses and the lowest achievable QBER in the HD-QKD session with Bob's measurement configuration.}
\label{tab1}
\end{center}
\end{table}

After the photon is transmitted through the MCF2 fiber it is detected at Bob's station for state analysis. Bob's detection scheme is similar to the one used by Alice. The output face of the MCF2 is magnified at a second deformable mirror (DM2) with a 10$\times$ objective lens (L4), and the relative-phase of each core is addressed individually by four independent mirrors. To define a common shared referential frame between the communicating parties, like in fiber-based polarization QKD schemes, Bob first defines the offset positions of the four mirrors for the post-selection of the state $|\varphi_1^{(1)}\rangle$, when Alice is also sending such state. Thus, compensating residual phase-shifts $\phi'_l$ acquired over the MCF2 propagation. By placing a ``pointlike'' single-photon detector at the DM2's far-field (FF) plane, and properly adjusting the mirrors longitudinal positions to set phase-shifts equal to $\pi$, Bob can post-select for detection any state $|\varphi_i^{(j)}\rangle$ required for the 4-dimensional BB84 QKD session. In our case the ``pointlike'' single-photon detector is composed of a pinhole (PH) fixed at the center of the FF plane of a lens L5 ($f_{L5}=7.5$~cm), a single-mode fiber, and an InGaAs avalanche single-photon detector (SPD) [See Fig.~\ref{Fig1}(a)]. The probability that a photon is detected at the center of the FF plane, $C_s$, is proportional to the overlap between the generated and the post-selected states [See details at \cite{Etcheverry_Scirep_2013, Glima_Opex_2011}]. At Fig.~\ref{Fig1}(d) we show three examples of the FF distribution, which arise from the phase modulations used by Alice and Bob to prepare and measure the states indicated below each figure. The red circle indicates the area of the employed pinhole. One can see that $C_s \propto |_{Alice}\langle\varphi_i^{(j)}|\varphi_{i'}^{(j')}\rangle_{Bob}|^2$, also with $i'=1,2,3,4$ and $j'=1,2$. Note that the pinhole diameter defines the overall quality of Bob's measurement, which in turn defines the losses and the lowest achievable QBER as we show in Table~\ref{tab1}. In our case we adopted the second configuration since lower error rates are preferable over losses for secret key bit generation rate. Last, the use of one single-photon detector configuration for the 4-dimensional QKD is discussed on the Methods section.

\subsubsection*{Fiber propagation, mean QKD state fidelity, referential frame control system}

\begin{figure}[t]
\centerline{\includegraphics[width=0.7 \textwidth]{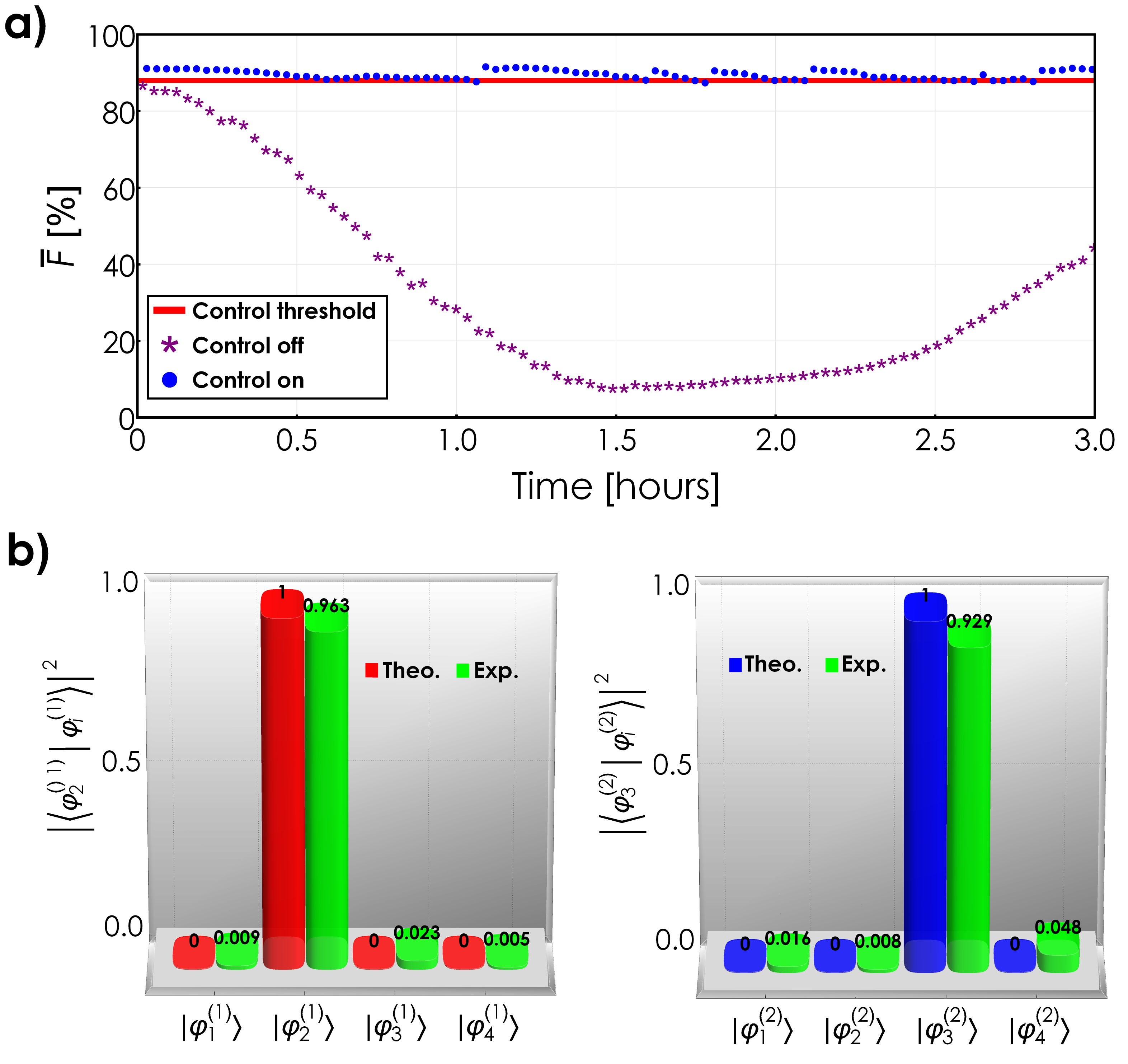}}
\caption{\textbf{Mean QKD state fidelity.} a) We show the mean QKD state fidelity, averaged over all the eight $4$-dimensional BB84 states transmitted through the 0.3~km long multicore fiber. In the case with the control off, the fidelity slowly degrades as a function of time due to the misalignment of Alice and Bob's shared referential frame. With the control turned on, the fidelity is always above the chosen threshold, thus enabling stable QKD sessions. b) Two examples of measured fidelities after 1.08 hours, for the states  $|\varphi_2^{(1)}\rangle$ and $|\varphi_3^{(2)}\rangle$. Theoretical and experimentally measured fidelities are shown.} \label{Fig2}
\end{figure}

The fact that the cores of multicore fibers lie within the same cladding make them intrinsically robust against random-phase fluctuations, as thermal and mechanical perturbations act globally over the core modes. Nonetheless for long multicore fibers, like the MCF2 used as our transmission channel, slowly time varying phase-drifts can still be observed. This effect deteriorates the referential frame shared by Alice and Bob, resulting in a mean QKD state fidelity ($\bar{F}\equiv 1-QBER$) that decreases over time as the error rate increases. The typical behaviour observed for $\bar{F}$ is shown with the purple star-dotted curve into Fig.~\ref{Fig2}(a). This renders HD-QKD over long/installed multicore fibers not practical if not properly addressed. To overcome this problem we developed a custom control system. It checks the referential frame of Alice and Bob stations over given time intervals of 30~$s$  and the control routine is initialized if the QBER surpasses a defined threshold value. During the control procedure the QKD session, which will be explained next, is interrupted. The control system is composed of a laser that is multiplexed into the multicore fibers, together with the attenuated pulses, and two field programmable gate arrays (FPGA1 and FPGA2) electronic modules that are used to actively control both deformable mirrors of the setup [See Fig.~\ref{Fig1}(a)]. Based on a custom designed closed-loop maximum-power-point-tracking algorithm, the control system varies the offset positions of all the active mirrors used on the QKD session until the recorded QBER is back to a value below our defined threshold of 12$\%$. Then, it is turned off and the QKD session restarts. Note that the control laser operates only during the referential frame control session, otherwise the security of our QKD session would be compromised. The resulting effect of the control system is shown into Fig.~\ref{Fig2}(a) with a blue dotted curve. One can see that it allows the stabilisation of the shared referential frame, critical for long-term QKD sessions. In Fig.~\ref{Fig2}(b) we show the fidelity measurement for the states $|\varphi_2^{(1)}\rangle$ and $|\varphi_3^{(2)}\rangle$ at 1.08 hours of test. The mean QKD state fidelity is $\bar{F}_{1.08}=(92.05\pm0.03)\%$ and the corresponding fidelity for each state is $(96.31\pm 0.03)\%$ and $(92.93\pm 0.03)\%$, respectively.

\subsubsection*{Which-path information erasure}

Before the QKD session is implemented, it is also important to consider that polarization drifts may occur over long multicore fibers. That is, different core modes can be associated to different polarization modes at the end of light propagation over the fiber. This would be a consequence of asymmetries of the transverse shape of the core modes arising from imperfections during the fabrication process, which may generate polarization mode dispersion with different intensities for each core. In this case, the polarization degree of freedom will partially mark the single-photon propagation path over the fiber, which in turn compromises the state coherence if the polarization is not also properly addressed by Bob. Fortunately, this effect can be fully compensated with the use of polarisation filters. In our case, the polarization-based distinguishability of the core modes was almost constant over time as our multicore fiber was protected inside the laboratory. Then, we used quarter-waveplates (QWP), half-waveplates (HWP) and polarizing beamsplitters (PBS) to erase the which-path information [See Fig.~\ref{Fig1}(a)]. The overall loss at the eraser stage was of only 1.2~dB. Note, however, that for installed multicore fibers active polarisation controls based on liquid crystals displays can be used.

\subsubsection*{Automated QKD session}

The 4-dimensional BB84 QKD session is also implemented by the two field programmable gate array electronic units. FPGA1, belonging to Alice, generates a 1 kHz synchronisation signal which is shared to FPGA2, owned by Bob. Following the generation of a sync pulse, FPGA1 reads a number from an idQuantique Quantis quantum random number generator (QRNG), which determines whether MUB $j=1$ or $j=2$ will be used, and which of the $|\varphi_i^{(j)}\rangle$ states from that MUB will be created at the DM1. After the active QKD mirrors from DM1 are ready for state generation, the attenuated optical pulse is created, and the state $|\varphi_i^{(j)}\rangle$ is encoded into a probabilistically generated photon, which is transmitted to Bob through MCF2. Bob's FPGA2 will have received the same sync pulse almost simultaneously as Alice generates it. He then also takes a number from his QRNG and chooses one of the two MUBs and one of the corresponding four states in which to project the incoming photon. This choice is forwarded to DM2, setting its QKD active mirrors into position, before the single photon is received by Bob. A delayed version of the synchronisation pulse is fed in the gated-mode single-photon detector (idQuantique id210), with the gate width adjusted to 0.85 ns. FPGA2 then checks whether there was a detection in the SPD for that particular sync pulse. Both FPGAs record in each measurement round, the chosen MUB and the corresponding state and whether a single-photon was detected. The FPGAs compare the detected strings after basis reconciliation to calculate the QBER.

\subsubsection*{QKD results}

\begin{figure}[t]
\centerline{\includegraphics[width=0.7\textwidth]{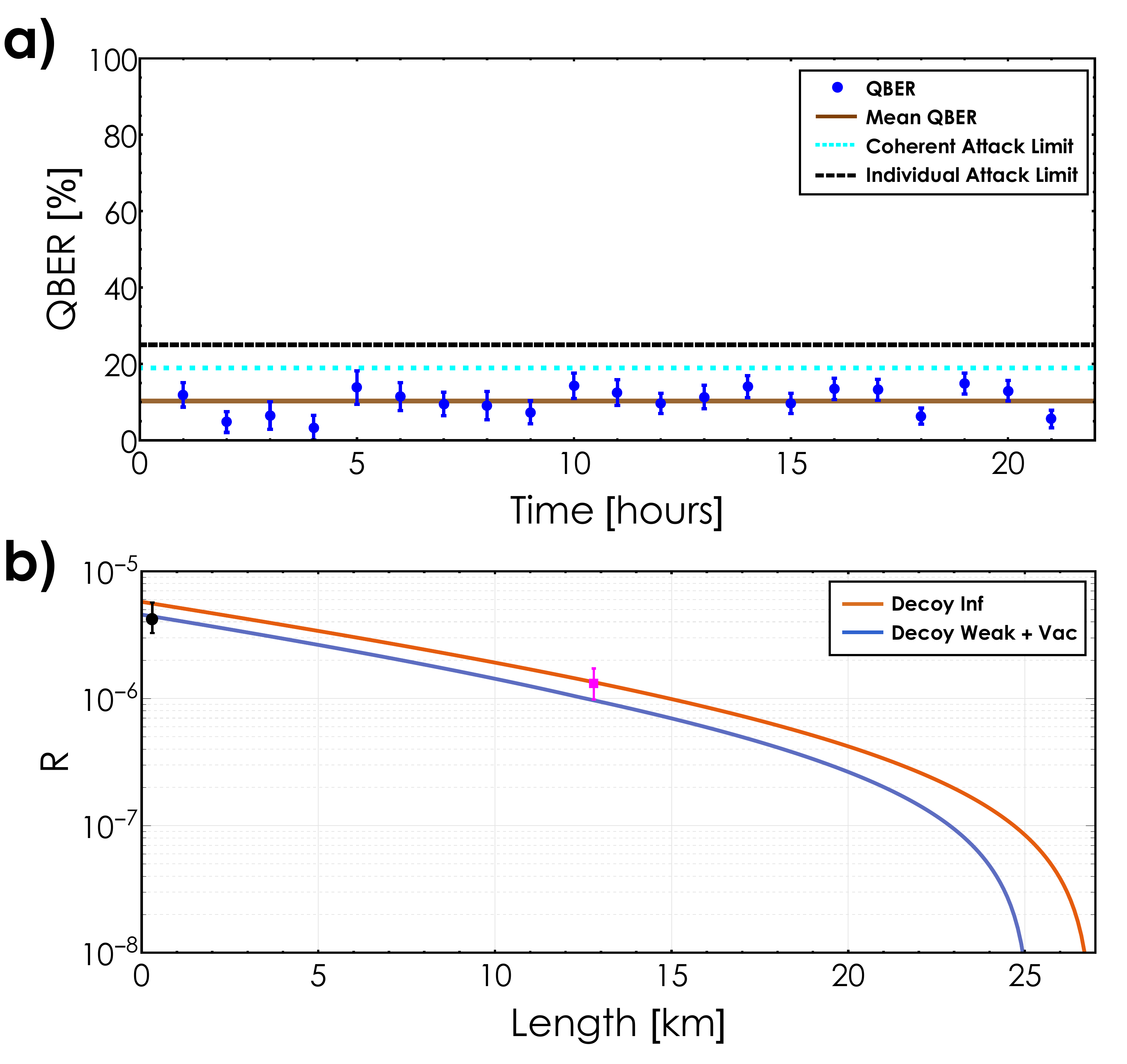}}
\caption{\textbf{Experimental QKD results.} a) Quantum bit error rate (QBER), as a function of time, with each data point indicating the average over the past hour. The brown line shows the average measured QBER of $10.25\%$ in a HD-QKD session with $\mu = 0.27$. The dashed black and cyan lines denote the theoretical upper bounds to achieve positive key rate for $d =4$, while considering individual ($25\%$) and coherent ($18.93\%$) attacks, respectively. b) Secret key generation rate (R) as a function of distance, while considering the upper bound case of infinite decoys (red curve) and the practical weak decoy + vacuum protocol (blue curve). The two data points correspond to the actual key generation rate for the QKD experimental runs performed (see text for details), for the weak decoy + vacuum (black circle) and infinite decoy (magenta square) cases. Error bars correspond to propagated errors arising from the Poissonian detection statistics.}\label{Fig3}
\end{figure}

The secret key generation rate (R) as a function of the dimension $d$ is given by (see Methods for details)
\begin{eqnarray}
R &\ge &Q_0  log_2 d + Q_1 \left[log_2 d - H_d \left(e_1\right)\right]-Q_\mu H_d \left(E_\mu \right)f\left(E_\mu\right),
\label{eq:rate}
\end{eqnarray} where $Q_0$ and $Q_1$ are the gains of the vacuum and single photon states, respectively. $Q_\mu$ is the experimentally measured gain for an average $\mu$ photon number. $H_d\left(x\right) = -x log_2 \left[x/\left(d-1\right)\right] - \left(1-x\right) log_2 \left(1-x\right)$ is the $d$-dimension modified Shannon entropy of the QBER, which considers that the error can randomly occur in any of the $d-1$ detectors \cite{Sheridan_PRA_2010}. $e_1$ is the single-photon error rate, $E_\mu$ is the measured overall
quantum bit error rate (QBER), and $f\left(E_\mu\right)$ is the inefficiency of the error correction function. We have employed $f\left(E_\mu\right)=1.05$ \cite{Elkouss_QIC_2011} since reported error rates in typical HD-QKD experiments, including ours, hover around 10\% \cite{Etcheverry_Scirep_2013, Boyd_NJP_2015}. The secret key probability considers the use of the efficient BB84 protocol \cite{Lo_JC_2005}, while an additional factor $1/d$ is required if all bases were employed with equal probability.

We first performed a long term automated measurement to demonstrate the stability achieved in our experiment by performing a BB84 QKD session over the 0.3 km of multicore fiber, while employing an average photon number per pulse $\mu = 0.27$. The results are shown in Fig.~\ref{Fig3}(a), where we have an average of 44.5 detections per hour, with an average QBER of $10.25\%$. The results clearly show that the control system is able to minimize phase drifts during the run, while keeping a QBER considerably lower than the security thresholds. The thresholds are $18.93\%$ and $25\%$ for collective and individual attacks, respectively \cite{Cerf_PRL_2002}.

Based on this measured mean $10.25\%$ QBER over the entire session, we estimated the key rate as a function of distance by optimizing over  $\mu$ and assuming the infinite decoy case (see Methods). The result is represented by the red curve in Fig. \ref{Fig3}(b), which gives an upper bound for the key generation rate in our system.

We also calculated the rate as a function of the fiber length by using the well known, and practical, vacuum + weak decoy protocol \cite{Ma_PRA_2005}. It consists of two weak decoy states (of which one is the vacuum) and a stronger signal state. Based on our experimental setup characteristics, we fix the weak decoy mean photon number per pulse to $\nu=0.1$ (a compromise between optimizing the key rate and the estimation of $e_1$ and $Q_1$) and optimize the mean photon number $\mu$ of the signal state to obtain the secret key rate. The result is given by the blue curve in Fig. \ref{Fig3}(b). This clearly shows we can generate positive secret key rates up to 25 km of multicore fiber when using a realistic decoy protocol with standard components and single-photon detectors. To demonstrate it, we performed the key exchange section while employing the value of $\mu=0.2$ for the signal at the distance of 0.3 km, and the decoy states ($\nu=0.1$ and vacuum). We finally obtain a secret key generation rate per pulse of $(4.31 \pm 1.19) \times 10^{-6}$, plotted as the black dot in Fig.~\ref{Fig3}(b). Table~\ref{tab2} displays the measured parameters that are used to calculate the key rate at the distance of 0.3 km. (for details please see Methods). Our calculation returns a lower bound for the single-photon gain $Q^L_1 = (6.96 \pm 1.30) \times 10^{-6}$ and an upper bound of the single-photon error rate $e_1^{U} = (7.53\pm2.20)\%$.

\begin{table*}[t]
\centering
\begin{tabular}{ | c | c | c|}
\hline
		 Signal ($\mu$)
		& Weak ($\nu$)
		& Vacuum \\
\hline

$Q_\mu=(9.31\pm0.63) \times 10^{-6}$
&$Q_\nu=(4.89\pm0.30) \times 10^{-6}$
& $Y_0=(2.06\pm0.23) \times 10^{-7}$\\
\hline

$E_\mu=(10.8 \pm 1.4)\%$
&$E_\nu=(9.0 \pm 1.3)\%$
&$E_0=(71.1 \pm 3.4)\%$ \\
\hline
\end{tabular}
\caption[caption]{\textbf{Measured parameters for the weak decoy + vacuum protocol.} Experimental results used to obtain the secret key generation rate at a distance of 0.3 km, when using the weak decoy + vacuum protocol.}\label{tab2}
\end{table*}

Lastly we performed a new HD-QKD run with an extra 5 dB optical attenuator placed before the detector to simulate a total transmission distance of 12.8 kms (as our multicore fiber is specified to have an attenuation coefficient of 0.4 dB/km), assuming an infinite number of decoy states. The goal is to demonstrate an upper bound for the rate at a longer distance. This was executed over a continuous period of 45.3 hours, with an average QBER of $9.80 \pm 1.69\%$, with a secret key generation rate per pulse of $(1.30 \pm 0.36) \times 10^{-6}$ (shown as the magenta square in Fig.~\ref{Fig3}b).
\subsubsection*{MCF mode sorter}

\begin{figure}[t]
\centerline{\includegraphics[width=0.7 \textwidth]{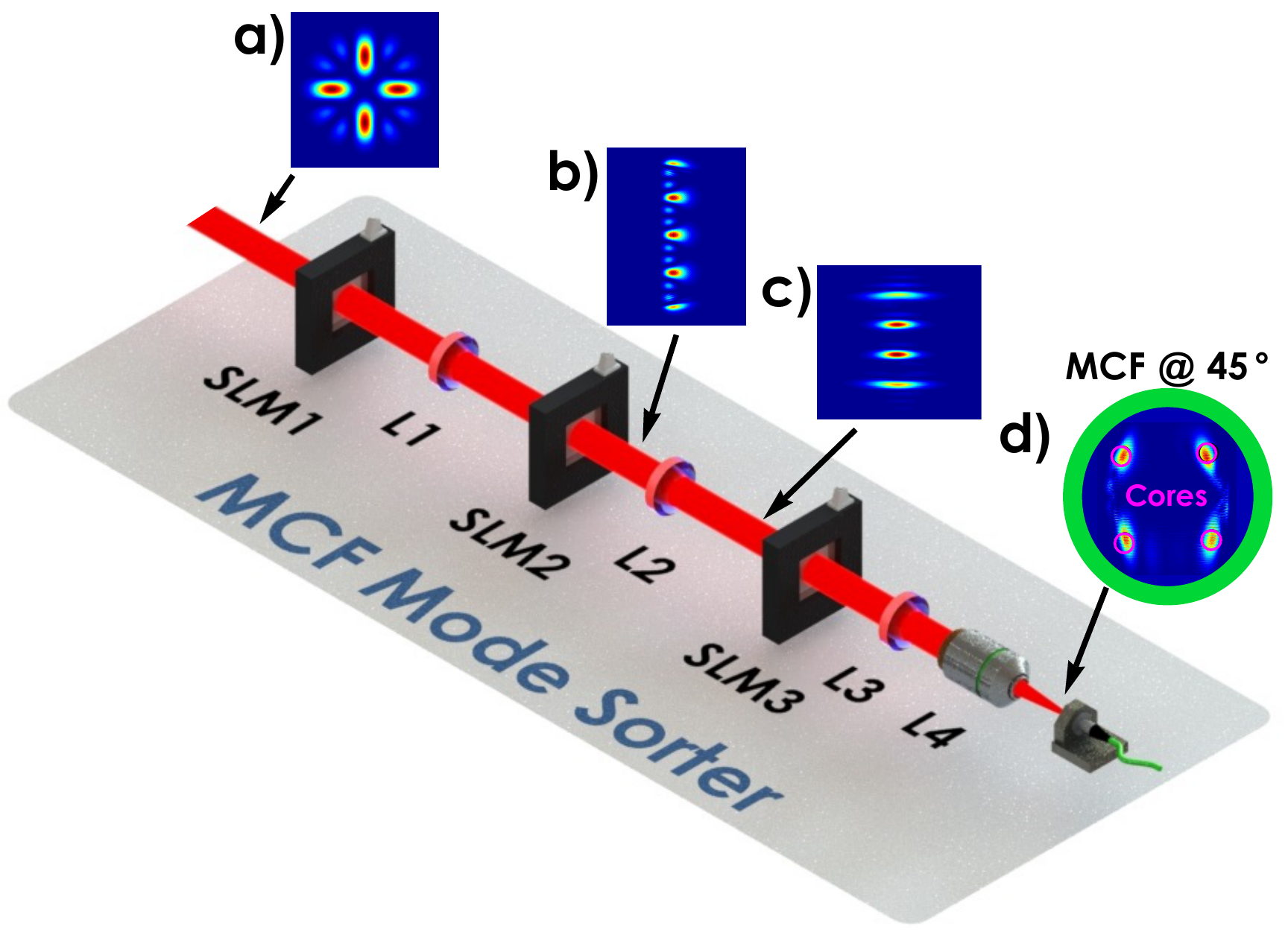}}
\vspace{-0.2cm}
\caption{\textbf{MCF mode sorter}. Proposed scheme of a mode sorter that maps OAM Laguerre-Gauss (LG) modes to transverse propagation modes (and vice-versa) compatible with the core geometry of a multicore fiber. The device is based on the use of spatial light modulators (SLM) and lenses to implement coordinate transformations. a) Input superposition of four OAM LG modes defined by the azimuthal index $l \in \{-6, -2, 2, 6 \}$. Then in b) and c), different transformations are applied separating the LG modes into different transverse modes (see main text for details). Finally in d), the mode sorting procedure is completed, with each OAM LG mode being coupled into a different core. An overall coupling efficiency of $40\%$ is attainable, while cross-talk mode coupling is negligible $< 0.15\%$.}
\label{Fig4}
\end{figure}

Another alternative for high dimensional quantum communication is based on free-space propagating OAM LG modes of light, which span a theoretically infinite discrete Hilbert space. Long distance OAM free-space classical communication have been demonstrated, suggesting the feasibility of practical quantum communications based on OAM through free-space \cite{Krenn_NJP_2014}. Here we present a device to create a flexible hybrid high-dimensional network, which can be used to interconnect multicore fiber based ground stations to OAM based free-space optical links. In this way, Alice would be able to communicate with Bob even when there is no line-of-sight between them, and also no deployed multicore fibers connecting them. For this purpose, she would encode information into high-dimensional OAM quantum states, send them through a free-space link, and at an intermediate station (sharing a line-of-sight with Alice) connected to the multicore optical network, Alice's states would be feed-forward to Bob's station. Our device is a modified version of a previously reported OAM mode sorter, which is capable of mapping different OAM modes into distinct transverse propagation modes \cite{Berkhout_PRL_2010}. Our modification relies on an extra new transformation that maps the already separated modes into the typical core mode configuration of multicore fibers. This scheme provides high phase stability and efficient coupling.

The proposed MCF mode sorter scheme is shown in Fig.~\ref{Fig4}. To demonstrate its viability by numerical simulations, we consider (without loss of generalization) that an equally weighed superposition of four Laguerre-Gauss modes defined by the azimuthal index $l \in \{-6, -2, 2, 6 \}$ is sent through the MCF mode sorter [See Fig.~\ref{Fig4}(a)]. The first spatial light modulator (SLM1) and a thin lens (L1) are used to perform a log-polar optical coordinate transformation by using the phase modulation $\Phi_1 = \frac{2 \pi a}{\lambda f_1} [ y_1 \arctan(y_1/x_1) - x_1 \ln (\sqrt{x_1^2+y_1^2}/b + x_1)]$ at the SLM1. The second SLM2 imprints the phase $\Phi_2 = - \frac{2 \pi a b}{\lambda f_1}\exp \left(- \frac{ x_2}{a}\right) \cos\left(\frac{y_2}{a}\right)$ to implement a required phase correction. The resulting transverse profile is given in Fig.~\ref{Fig4}(b). The second lens (L2) performs the mode separation exploiting the different phase gradients of each OAM mode [Fig.~\ref{Fig4}(c)]. The last components SLM3 and L3 perform a new transformation, that maps the linearly distributed modes to a configuration typical of multicore fibers (rotated by $45^{\circ}$). The phase necessary for this new transformation is $\Phi_3 =  \frac{2 \pi w}{\lambda f_3} (s x_3- \zeta) \sin(a_1 (y_3-\beta))$, where we have added correction parameters $s$, $\zeta$ and $\beta$ for fine grain adjustment \cite{Hossack_JMO_1987}. In our numerical simulation we used focal lengths  $f_1=180$ mm, $f_2 = 180 $ mm and $f_3 = 12$ mm, the wavelength $\lambda = 1550$ nm, and a beam waist $w_0 = 3200$ $\mu$m. The other parameters are: $a=0.001$, $b=0.007$, $c = 1.2$ m, $s = 0.8$, $\zeta = 300$ nm. $a_1 = \gamma \pi d/ [(l_4 + l_3) \lambda f_2]$ is the condition for fiber core matching, with the parameters $\gamma = 0.975$, $d= 2 w_0$, $\beta =b_1 \lambda l_3 /d$, $b_1 = 1.5$ m, where $l_4$, $l_3$ are the highest and the second azimuthal indexes $l$. Note that this transformation can be applied to other OAM mode superpositions. After being transmitted by a final $20 \times$ objective (L4), the light is coupled into the multicore fiber as it is shown in Fig.~\ref{Fig4}(d). The average coupling efficiency is $40\%$, and cross-talk effects are negligible $< 0.15\%$. That is, each OAM mode couples efficiently to a different core of the multicore fiber and the introduced error rate is negligible.

\section*{Discussion}

Quantum key distribution has been the most successful protocol of quantum communication, with many different demonstrations performed across several distinct scenarios. The interest on QKD is only expected to grow more given the recent developments aiming at the construction of a quantum computer and the demonstration of metropolitan QKD networks \cite{Lo_natphoton_2014}. Inline with standard communication systems some experiments have focused on increasing the transmission rate, which is arguably a major Achilles heel of QKD.

Considerable effort is being made to increase QC's information content by using the transverse spatial profile of a single-photon \cite{Steve_PRL_2006, Howell_PRL_2007, Etcheverry_Scirep_2013, Mafu_PRA_2013, Boyd_NJP_2015}. However, no approach so far has been proven to be fully compatible with a secure QKD session over the infrastructure already developed for classical telecommunications. In this work we show, for the first time, that stable and secure high-dimensional quantum communication is feasible over long-distance optical fibers. In our work 4-dimensional quantum states are encoded onto the linear-transverse momentum of single-photons and successfully transmitted over 0.3~km of a telecommunication multicore fiber. We demonstrated a fully automated and secure HD-QKD session by resorting to the decoy-state method. Our results set the stage for future implementations of high-dimensional quantum communication over the telecommunication infrastructure, constituting an important block of tomorrow's quantum internet. Our technique is also compatible with high-speed QKD links over long distances since, on the one hand, spatial light modulators with MHz repetition frequencies have been recently developed \cite{Qiu_scirep_2012}, while on the other hand the use of highly efficient superconducting detectors will greatly increase the maximum achievable distance \cite{Marsilli_Natphoton_2012}. Finally, it is shown that our work paves the way towards a high-dimensional quantum network composed of interconnected free-space and optical fiber links.

\section*{Methods}

\subsubsection*{The 4-dimensional BB84 QKD states used}

The 4-dimensional BB84 QKD session requires that Alice and Bob prepare eight states spanning two MUBs. These states will be denoted by $|\varphi_i^{(j)}\rangle$, where $i=1,2,3,4$ refers to the $i$th state of the $j$th MUB, with $j=1,2$. The states of the first MUB are defined by: $\langle\varphi_1^{(1)}|=\frac{1}{2}[1,1,1,1]$, $\langle\varphi_2^{(1)}|=\frac{1}{2}[1,-1,1,-1]$, $\langle\varphi_3^{(1)}|=\frac{1}{2}[1,1,-1,-1]$ and $\langle\varphi_4^{(1)}|=\frac{1}{2}[1,-1,-1,1]$. The second MUB states are: $\langle\varphi_1^{(2)}|=\frac{1}{2}[1,1,1,-1]$, $\langle\varphi_2^{(2)}|=\frac{1}{2}[1,1,-1,1]$, $\langle\varphi_3^{(2)}|=\frac{1}{2}[1,-1,1,1]$ and $\langle\varphi_4^{(2)}|=\frac{1}{2}[-1,1,1,1]$. The states are expressed in the basis of the four fiber core modes indicated in Fig.~\ref{Fig1}(b).

\subsubsection*{$d$ detectors vs. 1 detector scheme configuration}
\begin{figure}[t]
\centerline{\includegraphics[width=1 \textwidth]{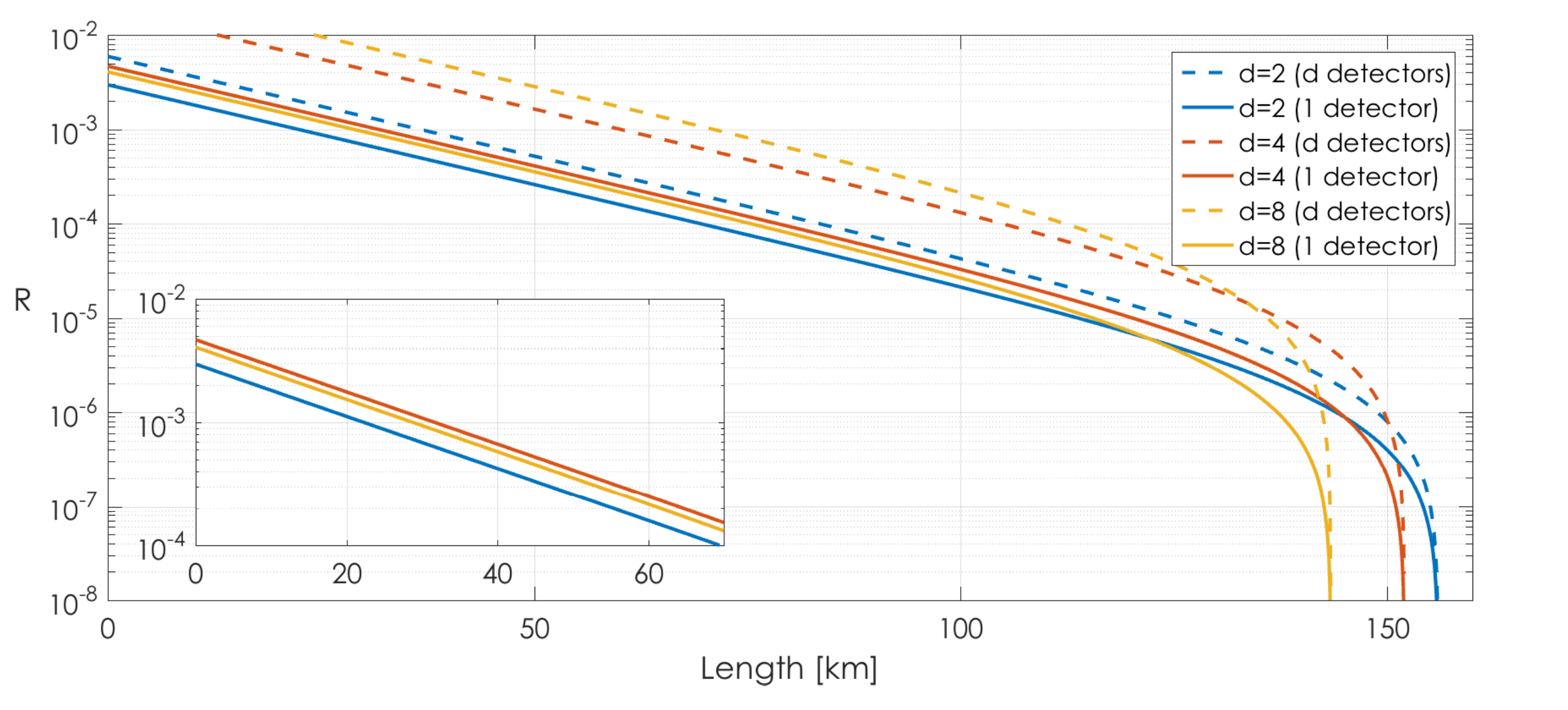}}
\caption{\textbf{Secret key rate for $d$ dimensional Hilbert spaces when considering a single-detector and a $d$-detector scheme configuration.} Here we perform secret key rate simulations considering the infinite decoy case, while using as input parameters the data from \cite{Gobby_APL_2004}. In the $d$-detector case, as expected, the rate increases for shorter distances and the cut-off point in the secret rate vs. distance curve occurs for shorter distances, as $d$ increases. In the single-detector case, the probability of correctly projecting the transmitted state onto itself decreases linearly with $d$, while the information gain per transmitted photon only increases with $O(log(d))$. Thus, the case of $d = 8$ always generates a lower secret key rate when compared to d = 4. Nonetheless, as shown on the inset, our implementation (i.e. the $d=4$ case) is capable of beating $d= 2$ case.}
\label{Fig5}
\end{figure}

Similarly to classical communications, the use of higher-dimensions to encode more information has a consequence: higher sensitivity to noise. In the case of a HD-QKD, there are two approaches: a single-detector randomly placed at each possible output (the one done in this work), or having $d$ detectors, one at each output. In the $d$-detector case, as expected, the rate increases for shorter distances and the cut-off point in the secret rate vs. distance curve occurs for shorter distances, as $d$ increases (Fig.~\ref{Fig5}). This is due to the fact that for each detection round, the total dark count probability is higher, compared to the two-dimensional case, as there are $d$ detectors. The single-detector case is more interesting, as can be also be seen in Fig.~\ref{Fig5}. In this case, the probability of correctly projecting the transmitted state onto itself decreases linearly with $d$, while the information gain per transmitted photon only increases with $O(log(d))$. Nonetheless, as shown on the inset of Fig.~\ref{Fig5}, our implementation (i.e. the $d=4$ case) is capable of beating $d= 2$ for all distances until the dark count probability gets too strong at the cut-off point.

\subsubsection*{Decoy-state secret key generation rate probability}

\begin{table}[t]
\centering
\renewcommand{\arraystretch}{1.4}
\begin{tabular}{| c | c | c | c | c |}
\hline
& $Q_\mu$
& $E_\mu$
& $Q_1$
& $e_1$
\\
\hline
$d$ detectors
&$Y_0+1-e^{-\mu\eta}$
&$\frac{e_0 Y_0+e_{opt} (1-e^{-\mu\eta})}{Y_0+1-e^{-\mu\eta}}$
&$(Y_0+\eta)\mu e^{-\mu}$
&$\frac{e_0 Y_0+e_{opt} \eta}{Y_0+\eta}$
\\
\hline
Single detector
&$Y_0+\frac{1-e^{-\mu\eta}}{d}$
&$\frac{e_0 Y_0 d+e_{opt} (1-e^{-\mu\eta})}{Y_0 d+1-e^{-\mu\eta}}$
&$(Y_0+\frac{\eta}{d})\mu e^{-\mu}$
&$\frac{e_0 Y_0 d+e_{opt} \eta}{Y_0 d+\eta}$
\\
\hline
\end{tabular}
\caption[caption]{System parameters for estimation of the secret key generation probability as a function of transmission distance.} \label{table:1}
\end{table}

Here we show how the secret key generation probability $R$ (namely the probability of obtaining a secure bit for each transmitted pulse) of a HD-QKD system can be derived using the decoy-state approach \cite{Hwang_PRL_2003, Lo_PRL_2005, Wang_PRL_2005}. Our analysis follows the method of Ref.~\cite{Ma_PRA_2005}, and modifications are performed when necessary for dealing with the high-dimensional case. We also show how the key rate is estimated as a function of the distance.

The secret key generation probability for a $d$-dimensional systems is given by \cite{Ma_PRA_2005, Lim_PRA_2014}
\begin{eqnarray}
R &\ge &Q_0  \log_2 d + Q_1 \left[\log_2 d - H_d (e_1)\right]-Q_\mu H_d (E_\mu) f(E_\mu)\,,
\label{eq:rate}
\end{eqnarray} where $Q_0$ and $Q_1$ are the gains of the vacuum and single-photon states, respectively.
$Q_\mu$ is the overall gain (i.e. the probability of obtaining a detection when the signal state is sent),
$E_\mu$ is the overall error rate, while $e_1$ is the error rate of the single-photon states. $H_d\left(x\right) = -x \log_2 \left[x/\left(d-1\right)\right] - \left(1-x\right) \log_2 \left(1-x\right)$ is the $d$-dimensional modified Shannon entropy of the QBER \cite{Sheridan_PRA_2010}; $f\left(E_\mu\right)$ is the inefficiency of the error correction function. The secret key probability considers the use of the efficient BB84 protocol \cite{Lo_JC_2005}.

The values of $Q_\mu$ and $E_\mu$ are directly obtained from the experimental data when Alice sends signal pulses. On the other hand, the parameters associated to single-photon pulses ($Q_1$ and $e_1$), and vacuum ($Q_0$), cannot be directly measured. They must be inferred through the use of an analytical or numerical approach based on the decoy-state technique \cite{Xu_PRA_2014}. A practical implementation of  consists on using only one weak (with average photon flux $\nu < \mu$) and vacuum decoy states. Under this approach, $Q_0$ can be directly estimated as $Q_0=e^{-\mu}Y_0$, where $Y_0$ is the measured
yield of the vacuum states (i.e. the probability of detection measured when no photons are sent from Alice).
On the other hand, a lower bound $Q^L_1$ on $Q_1$, and an upper bound $e^U_1$ of $e_1$, can be written as \cite{Ma_PRA_2005}
\begin{equation}
Q^L_1=\frac{\mu^2e^{-\mu}}{\mu\nu-\nu^2}\left[Q_\nu e^\nu - \frac{\nu^2}{\mu^2}Q_\mu e^\mu -\frac{\mu^2-\nu^2}{\mu^2}Y_0\right],
\end{equation} and
\begin{equation}
e^U_1 = \left(E_\nu Q_\nu \mu e^\nu - \mu e_0 Y_0\right)/\left(\nu Q^L_1 e^\mu\right),
\end{equation} with $Q_\nu$ and $E_\nu$ measured with the weak decoy state. These values are fed into Eq.~(\ref{eq:rate}) to calculate the experimental secret key rate.

The same method can be exploited to derive the expected key rate as a function of the channel length.
In this case the values of $Q_\mu$, $Q_\nu$, $E_\mu$ and $E_\nu$ can be estimated by assuming the propagation in a lossy channel. When using a photon source modelled as an incoherent mixture of Fock states, given by the Poisson distribution $P_n=\mu^n e^{-\mu}/n!$, the overall gain and QBER values are computed through the summation over all possible states. Thus, $Q_\mu = \sum_{n=0}^\infty Y_n P_n$ and $E_\mu=\left(1/Q_\mu\right) \sum_{n=0}^\infty e_n Y_n P_n$. In the above expression
$Y_n$ is the $n$-photon yield, defined as the probability of detection at Bob's station when Alice sends an $n$-photon Fock state and $e_n$ is the corresponding error. The $n$-photon gain, $Q_n=Y_n P_n$, results from the product of the yield $Y_n$ and the probability $P_n$ of the state being produced by Alice.

In a lossy channel the expected value of $Y_n$ is $Y_n\approx Y_0+\eta_n$, where $Y_0$ is the vacuum yield -- related to the dark count probability of the SPD ($P_{dark}$). The parameter $\eta_n=1-\left(1-\eta\right)^n$ is related to the overall efficiency $\eta$ of the channel -- given by the detector efficiency and the internal transmittance of Bob's apparatus ($\eta_{SPD}=6.09\%$ and $\eta_{Bob}=24.5$ dB respectively, with the numerical values corresponding to our setup). The link transmittance is given by $10^{-\alpha L/10}$, with the attenuation coefficient represented by $\alpha$ [dB/km] and the transmission link length given by $L$ [km]. In our case, the multicore fiber used (Fibercore) has $\alpha=0.4$ dB/km. The error associated to the $n$-photon states can be estimated to be $e_n=\left(e_0 Y_0 + e_{opt} \eta_n \right)/Y_n$, where $e_{opt}$ is due to the optical misalignment of the detection system and is estimated to be $e_{opt}=(9.64\pm0.98)\%$ in our case.

In a $d$-dimensional QKD system employing $d$ outputs (one single-photon detector at each output), the yield of the vacuum states is $Y_0=1-\left(1-P_{dark}\right)^d$ which, for small values of $P_{dark}$, increases linearly with the dimension $Y_0 \approx d P_{dark}$. The QBER associated to vacuum states is $e_0=\left(d-1\right)/d$, corresponding to the probability of a random dark count to occur in an SPD which is not expected to fire when Alice and Bob's bases are matched.

With one single-photon detector in the $d$-dimensional case, the vacuum yield is independent of the dimension and limited to $Y_0=P_{dark}$. On the other hand, some non-vacuum states sent by Alice will not be measured by Bob, even in the case of compatible bases between Alice and Bob, and the overall efficiency is reduced to $\eta_n=\left[1-\left(1-\eta\right)^n\right]/d$.

The expected values of $Q_\mu$ and $E_\mu$ and the parameters associated to single-photon events, $Q_1$ and $e_1$
for a given overall channel efficiency $\eta$ and a $d$-dimensional QKD system, are summarized in Table \ref{table:1} for both single and $d$ detector cases.
The curves for the secret key rate, as a function of the fiber length, shown on Fig.~\ref{Fig3} and Fig.~\ref{Fig5} are computed by feeding the values of table \ref{table:1} into Eq.~\eqref{eq:rate}.

\subsection*{Acknowledgments}

The authors thank Jan-\AA ke~Larsson for valuable discussions. This work was supported by Fondecyt~1160400, Fondecyt~1150101, CONICYT~PFB08-024,and Milenio~RC130001. G. C. acknowledges support from Fondecyt~11150324. E. S. G. acknowledges support from Fondecyt~11150325. P. W. R. C. acknowledges University of Birmingham Careers Network. A. P. acknowledges NCN Grant No. 2014/14/M/ST2/00818. N. V., J. C., P. G. acknowledge CONICYT.

\subsection*{Authors' Contributions}

G. B. X. and G. L. conceived the project. G. C., N. V., J. Cari\~ne, P. G., P. W. R. C. and E. S. G. built the experimental setup and performed the measurements with assistance from  G. B. X. and G. L. The referential frame control system was developed by G. C., N. V., J. Cari\~ne, J. Cardenas, M. F., G. B. X. and G. L. The theory and simulations for the new mode sorter were performed by N. V., A. P. and G. L. The theoretical work for the decoy-state analysis was developed by G. V., P. V. and T. F. S. The paper was written by G. C., N. V., G. B. X. and G. L. with assistance from G. V., P. V. and T. F. S.

\subsection*{Additional information}

Correspondence and requests for materials should be addressed to Gustavo Lima.
The authors declare no competing financial interests.

\end{document}